\documentclass[aps,prb,twocolumn,showpacs]{revtex4}%
\usepackage{graphicx}

\begin{document}


\title{Geometrical edge barriers and magnetization 
in superconducting strips with slits}

\author{Yasunori Mawatari}
\affiliation{%
   Energy Electronics Institute, 
   National Institute of Advanced Industrial Science and Technology,\\
   Tsukuba, Ibaraki 305--8568, Japan }
\author{John R.\ Clem}
\affiliation{%
   Ames Laboratory and Department of Physics and Astronomy, \\
   Iowa State University, Ames, Iowa, 50011--3160}

\date{February 6, 2003}

\begin{abstract}
We theoretically investigate the magnetic-field and current distributions for 
coplanar superconducting strips with slits in an applied magnetic field $H_a$. 
We consider ideal strips with no bulk pinning and calculate the hysteretic 
behavior of the magnetic moment $m_y$ as a function of $H_a$ due solely to 
geometrical edge barriers. We find that the $m_y$-$H_a$ curves are strongly 
affected by the slits. In an ascending field, the $m_y$-$H_a$ curves exhibit 
kink or peak structures, because the slits prevent penetration of magnetic flux. 
In a descending field, $m_y$ becomes positive, because magnetic flux is trapped 
in the slits, in contrast to the behavior of a single strip without slits, for 
which $m_y\approx 0$. 
\end{abstract}

\maketitle

\section{Introduction} 
Superconducting flat strips subjected to a perpendicular magnetic field show 
magnetic hysteresis, even when the strips have no bulk pinning. The magnetic 
hysteresis of strips without bulk pinning arises from geometrical edge 
barriers, i.e., barriers for magnetic-flux penetration at the strip 
edges.~\cite{Indenbom94,Shuster94,Zeldov94a,Benkraouda96,Maksimov98,Doyle97} 
Current-carrying strips have finite critical currents arising from the edge 
barriers,~\cite{Benkraouda98} and the critical current becomes larger when 
slits are fabricated near the edges of the strips.~\cite{Mawatari01} 
The critical-current increase is due to the enhancement of edge-barrier 
effects; in other words, making slits increases the number of edges that 
prevent flux penetration into the inner strips. The reversible magnetic 
response of two strips (i.e., a strip with a slit) in the Meissner state was 
considered in Refs.~\onlinecite{Zhelezina02} and \onlinecite{Brojeny02}. 
When an applied magnetic field exceeds a certain value, magnetic flux 
penetrates into the strips and the magnetic response becomes irreversible and 
hysteretic.~\cite{Zhelezina02} 

In this paper, we present a theoretical investigation of the magnetic 
hysteresis of bulk-pinning-free strips with slits in the presence of an 
applied magnetic field $H_a$. The hysteretic behavior of the magnetic moment 
$m_y$ as a function of $H_a$ shown in this paper is due solely to the 
geometrical edge barriers. In Sec.\ II we outline our theoretical approach and 
establish notation. In Sec.\ \ref{1_strip} we briefly review published results 
of the $m_y$-$H_a$ curves of a single strip without slits. In Sec.\ 
\ref{2_strips} we investigate field distributions and $m_y$-$H_a$ curves of two 
strips (i.e., a strip with a slit), and in Sec.\ \ref{3_strips} we study three 
strips (i.e., a strip with two slits). We briefly summarize our results in 
Sec.\ VI. 

\section{Complex field and magnetic moment} 
\begin{figure}
\includegraphics{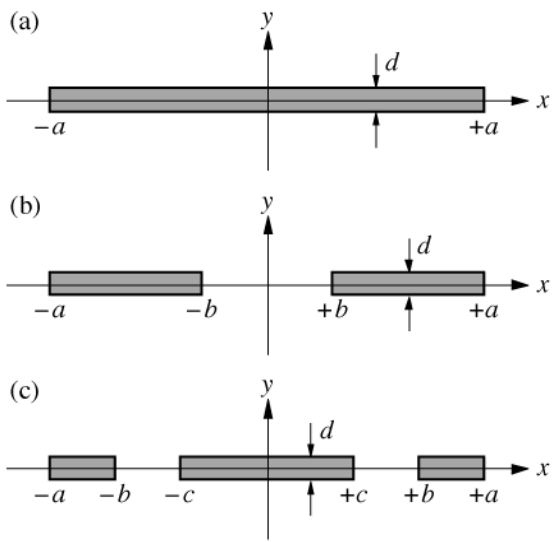}
\caption{%
Cross sections of strips with total width $2a$ and thickness $d$: 
(a) single strip without slit, (b) two strips (i.e., strip with a slit), 
and (c) three strips (i.e., strip with two slits).}
\label{Fig_strips}
\end{figure}

We investigate coplanar superconducting strips (i.e., strips in which slits are 
fabricated parallel to the edges) in a perpendicular magnetic field but 
carrying no net transport current. The strips under consideration have total 
width $2a$, thickness $d\ll 2a$, and infinite length along the $z$ axis (i.e., 
$|x|<a$ and $|y|<d/2$), as shown in Fig.~\ref{Fig_strips}. We assume that 
magnetic flux penetrates and escapes along the $x$ axis, assuming that there is 
no flux penetration from the ends at $|z|\to\infty$. (We may think of the strip 
ends as being connected by superconducting shunts.) 

The Biot-Savart law for the complex 
field~\cite{Zeldov94a,Clem73,Mawatari01,Brojeny02} 
${\cal H}(\zeta)=H_y(x,y)+iH_x(x,y)$ as a function of $\zeta=x+iy$ in the 
thin-strip limit, $d/a\to 0$, is expressed as 
\begin{equation}
   {\cal H}(\zeta)= H_a+\frac{1}{2\pi} 
     \int_{-a}^{+a}du \frac{K_z(u)}{\zeta-u} , 
\label{H-J}
\end{equation}
where the magnetic field $H_a$ is applied parallel to the $y$ 
axis, and $K_z(x)=\int_{-d/2}^{+d/2}j_z(x,y)dy$ is the sheet current along the 
$z$ axis. In Secs.\ \ref{2_strips} and \ref{3_strips} we show distributions of 
magnetic field $H_y(x,0)=\mbox{Re}[{\cal H}(x)]$ and current $K_z(x)/2 
=\mp H_x(x,\pm 0)=\mp\mbox{Im}[{\cal H}(x\pm i0)]$. The multipole expansion of 
Eq.~(\ref{H-J}) for $|\zeta|/a\to\infty$ is expressed as 
\begin{eqnarray}
   {\cal H}(\zeta) &\to& H_a+\frac{1}{2\pi} \int_{-a}^{+a}du K_z(u) 
     \left(\frac{1}{\zeta} +\frac{u}{\zeta^2} +\cdots\right) 
\nonumber\\
   &\to& H_a +\frac{I_z}{2\pi}\frac{1}{\zeta} 
     -\frac{m_y}{2\pi}\frac{1}{\zeta^2} +\cdots , 
\label{HJ-expand}
\end{eqnarray}
where $I_z=\int_{-a}^{+a}dxK_z(x)$ is the transport current 
along the $z$ axis and $m_y=\int_{-a}^{+a}dx(-x)K_z(x)$ is the magnetic moment 
in the $y$ direction per unit length. In this paper we consider the hysteretic 
relationship between $m_y$ and $H_a$ of strips carrying no net current 
($I_z=0$). 

The complex field for symmetrically arranged strips has the general form 
[e.g., Eqs.~(\ref{1-H}), (\ref{2-H}), and (\ref{3-H})] 
\begin{equation}
   {\cal H}(\zeta)=H_a\sqrt{\prod_n 
   \frac{\zeta^2-\alpha_n^2}{\zeta^2-a_n^2}} , 
\label{H_general}
\end{equation}
where the strip edges are at $x=\pm a_n$. 
Equation~(\ref{H_general}) may be expanded as 
\begin{equation}
   {\cal H}(\zeta)\to H_a +\frac{H_a}{2\zeta^2} 
     \sum_n \left(a_n^2-\alpha_n^2\right) +\cdots . 
\label{H_general_expand}
\end{equation}
Comparing Eq.~(\ref{H_general_expand}) with Eq.~(\ref{HJ-expand}), we obtain 
a general expression for the magnetic moment per unit length [e.g., 
Eqs.~(\ref{1-my}), (\ref{2-my}), and (\ref{3-my})], 
\begin{equation}
   m_y= -\pi H_a\sum_n\left( a_n^2-\alpha_n^2 \right) . 
\label{my_general}
\end{equation}

When $H_a$ is small enough, such that the local magnetic fields at the strip 
edges are all less than the flux-entry field 
$H_s$,~\cite{Benkraouda96,Benkraouda98,Mawatari01} the superconducting strips 
are in the Meissner state and no magnetic flux penetrates into the strips. 
References~\onlinecite{Zhelezina02} and \onlinecite{Brojeny02} describe the 
linear reversible magnetic response of two strips in the Meissner state. 
On the other hand, when $H_a$ is sufficiently large to make the edge fields 
reach $H_s$, magnetic flux penetrates into the strips, and the magnetic 
response becomes irreversible and hysteretic. In Secs.\ 
\ref{1_strip}-\ref{3_strips} we determine the parameters $\alpha_n$ and the 
magnetic moment $m_y$ as functions of $H_a$ in ascending ($H_{a\uparrow}$) and 
descending ($H_{a\downarrow}$) fields. 

\section{\label{1_strip}%
Single strip without slits} 
\subsection{Complex field for a single strip} 
In this section we consider a single strip of width $2a$, as shown in 
Fig.~\ref{Fig_strips}(a). The complex field ${\cal H}(\zeta)$ and the 
magnetic moment per unit length $m_y$ for a single strip are given 
by~\cite{Benkraouda96}
\begin{eqnarray}
   {\cal H}(\zeta) &=& H_a\sqrt{\frac{\zeta^2-\alpha^2}{\zeta^2-a^2}} , 
\label{1-H}\\
   m_y &=& -\pi H_a\left( a^2-\alpha^2 \right) . 
\label{1-my}
\end{eqnarray}

For convenience we introduce $a_{\pm}=a\pm\delta$, where $\delta$ is a cutoff 
length on the order of the thickness $d$ when the penetration depth 
$\lambda$ is small ($\lambda < d$) or the two-dimensional screening 
length $\Lambda = 2 \lambda^2/d$ when $\lambda$ is large ($\lambda > 
d$); i.e., $\delta \sim\max(d,\Lambda)$. 

\subsection{Single strip in an ascending field} 
\subsubsection*{Step (0) for $0<H_{a\uparrow}<H_0$} 
When a magnetic field $H_a$ is increased after zero-field cooling, the 
superconducting strips are initially in the Meissner state. The magnetic field 
at the edges is less than the flux-entry field $H_s$ [i.e., $H_y(a_+,0)<H_s$], 
and the edge barriers near $x\simeq\pm a$ prevent penetration of magnetic flux. 
In this step (0), the parameter in Eq.~(\ref{1-H}) is $\alpha=0$, and the 
magnetic moment is given by 
\begin{equation}
   m_y/\pi= -H_aa^2 . 
\label{1-m0}
\end{equation}
Magnetic flux cannot penetrate into the strips so long as 
$H_y(a_+,0)<H_s$ for $H_a<H_0$, but step (0) terminates when $H_y(a_+,0)=H_s$ 
at $H_a=H_0$, 
\begin{equation}
   \frac{H_0}{H_s}= \frac{\sqrt{a_+^2-a^2}}{a_+} 
   \simeq\sqrt{\frac{2\delta}{a}} , 
\label{1-H0}
\end{equation}
where the second equality is valid for $\delta/a\ll 1$. 

\subsubsection*{Step (i) for $H_0<H_{a\uparrow}<H_{\rm irr}$} 
When $H_a > H_0$, magnetic flux nucleates at $x\simeq \pm a$ and penetrates 
into the strip. A domelike distribution of magnetic flux exists for 
$|x|<\alpha_1$ and grows as $H_a$ increases. The parameter $\alpha=\alpha_1$ 
in Eq.~(\ref{1-H}) is determined by $H_y(a_+,0)=H_s$, 
\begin{eqnarray}
   \alpha_1^2 &=& a_+^2 -(H_s/H_a)^2 (a_+^2-a^2) 
\nonumber\\
   &\simeq& a^2-2\delta a[(H_s/H_a)^2-1]. 
\label{1-a1}
\end{eqnarray}
The magnetic moment is given by 
\begin{equation}
   m_y/\pi= -H_a(a^2-\alpha_1^2) 
     \simeq 2(H_a -H_s^2/H_a)a\delta . 
\label{1-m1}
\end{equation}

Step (i) terminates when $\alpha_1=a_-$ at $H_a=H_{\rm irr}$, where 
\begin{equation}
   \frac{H_{\rm irr}}{H_s} 
   = \sqrt{\frac{a_+^2-a^2}{a_+^2-a_-^2}} 
   = \sqrt{\frac{2a+\delta}{4a}} 
   \simeq \frac{1}{\sqrt{2}} . 
\label{1-Hirr}
\end{equation}

\subsubsection*{Step (ii) for $H_a>H_{\rm irr}$} 
The domelike distribution of magnetic flux expands to include almost all of 
the strip, and the strip's magnetic response is reversible for 
$H_a>H_{\rm irr}$. The edge field at $H_a=H_{\rm irr}$ is given by 
$H_y(a_+,0)\simeq H_s$. 

\subsection{Single strip in a descending field} 
\subsubsection*{Step (iii) for $0<H_{a\downarrow}<H_{\rm irr}$} 
In a descending field, magnetic flux escapes from the strip, but a domelike 
distribution of magnetic flux remains in the strip. The detailed behavior of 
the field distributions and $m_y$ in descending fields depends upon the 
treatment of edges of a strip.~\cite{Zeldov94a,Benkraouda96,Doyle97} 
Here we adopt a simple model and put $\alpha=a_-$, which results in a domelike 
distribution of magnetic flux for $|x|<a_-$ and a small but sharply peaked 
current density flowing in the vicinity of the edges, $a_-<|x|<a$. 
The magnetic moment is given by 
\begin{equation}
   m_y/\pi= -H_a(a^2-a_-^2)\simeq -2H_a a\delta , 
\label{1-m2}
\end{equation}
which qualitatively agrees with that predicted by more detailed 
investigations.~\cite{Zeldov94a,Benkraouda96,Doyle97} 

Note that $m_y$ takes a very small negative value (i.e., $0<-m_y\ll H_s a^2$), 
because a single strip cannot trap any magnetic flux in a descending field. 
Geometric edge barriers in a single strip without slits do not prevent escape 
of magnetic flux. At $H_a=0$ magnetic flux is entirely removed from a single 
strip, resulting in zero remanent magnetic moment, $m_y=0$ at 
$H_{a\downarrow}=0$. For strips with slits, on the other hand, magnetic flux 
is trapped in the slits, and $m_y$ becomes positive in descending fields, 
as we show in Secs.\ \ref{2_strips} and \ref{3_strips}. 

\section{\label{2_strips}%
Two strips (strips with a slit)} 
\subsection{Complex field for two strips} 
In this section we consider two strips (i.e., a strip with a single slit) of 
total width $2a$, as shown in Fig.~\ref{Fig_strips}(b). The superconducting 
strips are at $b<|x|<a$, and the slit is centered between the strips, $|x|<b$. 
We assume that the two strips are connected at the ends ($|z|\to\infty$) by 
superconducting shunts, such that an applied magnetic field $H_a$ induces a 
circulating current in the two strips, $\int_{-a}^{-b}K_zdx 
=-\int_{+b}^{+a}K_zdx\neq 0$. General expressions for the complex field and 
the magnetic moment per unit length for two strips are given 
by~\cite{Brojeny02}
\begin{eqnarray}
   {\cal H}(\zeta) &=& H_a\sqrt{\frac{(\zeta^2-\alpha^2) 
     (\zeta^2-\beta^2)}{(\zeta^2-a^2)(\zeta^2-b^2)}} , 
\label{2-H}\\
   m_y &=& -\pi H_a\left( a^2+b^2-\alpha^2-\beta^2 \right) . 
\label{2-my}
\end{eqnarray}

In this section we use parameters $a_{\pm}=a\pm\delta$ and $b_{\pm} 
=b\pm\delta$ [where $\delta \sim\max(d,\Lambda)$], and define the 
function $f_2(x)=(x^2-a^2)(x^2-b^2)$ for convenience.

\begin{figure*}
\includegraphics{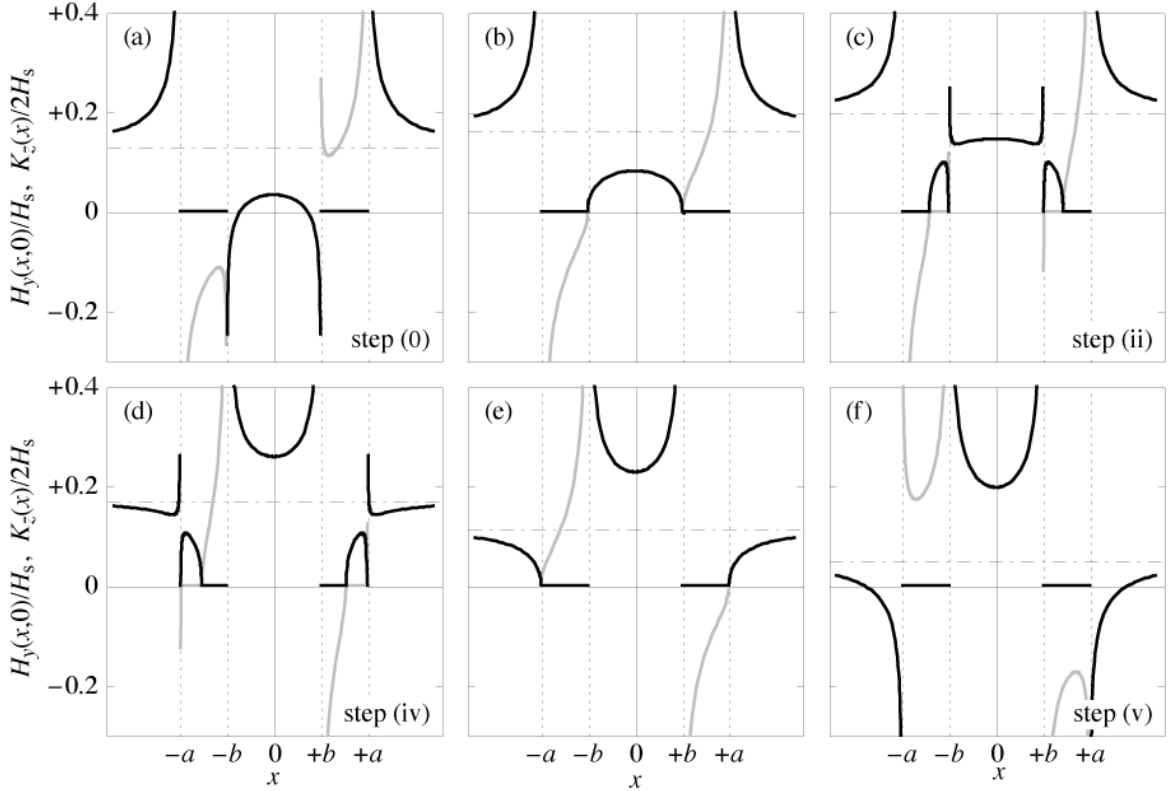}
\caption{%
Field distributions of $H_y(x,0)$, shown as black lines, and $K_z(x)/2=\mp 
H_x(x,\pm 0)$, shown as gray lines, in two strips for which $b/a=0.5$: 
(a) $0<H_{a\uparrow}<H_0$, (b) $H_{a\uparrow}=H_1$, 
(c) $H_1<H_{a\uparrow}<H_{\rm irr}$, (d) $H_4<H_{a\downarrow}<H_{\rm irr}$, 
(e) $H_{a\downarrow}=H_4$, and (f) $H_5<H_{a\downarrow}<H_4$. 
The horizontal dot-dashed lines show the applied magnetic field $H_a$.}
\label{Fig_HJ-2}
\end{figure*}

\subsection{Two strips in an ascending field} 
\subsubsection*{Step (0) for $0<H_{a\uparrow}<H_0$} 
When a magnetic field $H_a$ is applied after zero-field cooling, the 
superconducting strips are initially in the Meissner state. The magnetic 
fields at the edges are less than the flux-entry field, $0<-H_y(b_-,0) 
<H_y(a_+,0)<H_s$, and no magnetic flux penetrates into the strips. 
Figure~\ref{Fig_HJ-2}(a) shows distributions of the perpendicular field 
$H_y(x,0)$ and current $K_z(x)/2=\mp H_x(x,\pm 0)$ in step (0). The 
parameters in Eq.~(\ref{2-H}) are given by $\alpha=\beta=\alpha_0<b$. 
Because the total magnetic flux in the slit is zero, $\int_{-b}^{+b} 
H_y(x,0)dx=0$, the parameter $\alpha_0$ is determined as 
\begin{equation}
   \alpha_0^2=a^2 \left[1-E(b/a)/K(b/a)\right] , 
\label{1-a0}
\end{equation}
where $K(k)$ and $E(k)$ are the complete elliptic integrals of the 1st and 
2nd kind, respectively. The magnetic moment is given by 
\begin{equation}
   m_y/\pi= -H_a(a^2+b^2-2\alpha_0^2) . 
\label{2-m0}
\end{equation}

Step (0) terminates when $H_y(a_+,0)=H_s$ at $H_a=H_0$, where 
\begin{equation}
   \frac{H_0}{H_s}= \frac{\sqrt{f_2(a_+)}}{a_+^2-\alpha_0^2} 
   \simeq \frac{\sqrt{2\delta a(a^2-b^2)}}{a^2-\alpha_0^2} . 
\label{2-H0}
\end{equation}

\subsubsection*{Step (i) for $H_0<H_{a\uparrow}<H_1$} 
During this step, magnetic flux nucleates at the outer edges ($x=\pm a$), 
flows inward across the strips, and penetrates into the slit. Thus, we have 
$H_y(a_+,0)=H_s$ and $\int_{-b}^{+b} H_y(x,0)dx>0$. The field distributions 
in step (i) are similar to those in Fig.~\ref{Fig_HJ-2}(a). The parameter 
$\alpha=\beta=\alpha_1$ is determined by $H_y(a_+,0)=H_s$, 
\begin{eqnarray}
   \alpha_1^2&=& a_+^2 -(H_s/H_a) \sqrt{f_2(a_+)} 
\nonumber\\
   &\simeq& a^2-(H_s/H_a)\sqrt{2\delta a(a^2-b^2)} . 
\label{2-a1}
\end{eqnarray}
The magnetic moment $m_y$ is given by 
\begin{eqnarray}
   m_y/\pi &=& -H_a(a^2+b^2-2\alpha_1^2) 
\nonumber\\
   &\simeq& H_a(a^2-b^2) -2H_s\sqrt{2\delta a(a^2-b^2)} . 
\label{2-m1}
\end{eqnarray}

Step (i) terminates when $\alpha_1=b_+$ at $H_a=H_1$, where 
\begin{equation}
   \frac{H_1}{H_s} 
   = \frac{\sqrt{f_2(a_+)}}{a_+^2-b_+^2} 
   \simeq \sqrt{\frac{2\delta a}{a^2-b^2}} . 
\label{2-H1}
\end{equation}
The field distribution at $H_a=H_1$ is shown in Fig.~\ref{Fig_HJ-2}(b).

\subsubsection*{Step (ii) for $H_1<H_{a\uparrow}<H_{\rm irr}$} 
During this step, magnetic flux penetrates from $x=\pm a$, because 
$H_y(a_+,0)=H_s$. Domelike distributions of magnetic flux appear in the 
strips at $b_+<|x|<\alpha_2$ [Fig.~\ref{Fig_HJ-2}(c)], and grow as $H_a$ 
increases. Magnetic flux simultaneously exits the superconducting strips 
at $x= \pm b$ and enters the slit. Current spikes occur near $x\simeq\pm b$ 
in Fig.~\ref{Fig_HJ-2}(c) because a finite current $K_z\neq 0$ flows in the 
vicinity of the inner edges, $b<|x|<b_+$, whereas $K_z=0$ for $b+<|x|< 
\alpha_2$. In our theory, such current spikes, which produce wiggles in the 
local magnetic field distribution, always occur where magnetic flux is 
exiting from a dome in a superconducting strip. 

The parameters are given by $\alpha=\alpha_2$ and $\beta=b_+$, where 
$\alpha_2$ is determined by $H_y(a_+,0)=H_s$, 
\begin{eqnarray}
   \alpha_2^2 &=& a_+^2 -\left(\frac{H_s}{H_a}\right)^2 
     \frac{f_2(a_+)}{a_+^2-b_+^2} 
\nonumber\\
   &\simeq& a^2-2\delta a[(H_s/H_a)^2-1] . 
\label{2-a2}
\end{eqnarray}
The magnetic moment $m_y$ is given by 
\begin{eqnarray}
   m_y/\pi &=& -H_a (a^2+b^2-\alpha_2^2-b_+^2) 
\nonumber\\
   &\simeq& -2(H_s^2/H_a)a\delta +2H_a(a+b)\delta . 
\label{2-m2}
\end{eqnarray}

Step (ii) terminates when $\alpha_2=a_-$ at the irreversibility field 
$H_a= H_{\rm irr}\simeq H_s/\sqrt{2}$, when domelike flux distributions 
essentially fill the strips. The domes occupy the regions $b_+<|x|<a_-$. 

\subsubsection*{Step (iii) for $H_a > H_{\rm irr}$} 
When $H_a=H_{\rm irr}$, domelike flux distributions fill the regions 
$b_+<|x|<a_-$, and the magnetic response becomes reversible. At $H_a\simeq 
H_{\rm irr}$, the magnetic fields at the strip edges are $H_y(a_+,0)\simeq 
H_y(b_-,0)\simeq H_s$, and the parameters are given by $\alpha=a_-$ and 
$\beta=b_+$. The magnetic moment $m_y$, which is due to currents flowing in 
the narrow regions near the edges ($b < |x| < b_+$ and $a_- < |x| < a$), 
is given by 
\begin{eqnarray}
   m_y/\pi &=& -H_{\rm irr} (a^2+b^2-a_-^2-b_+^2) 
\nonumber\\
   &\simeq& -2H_{\rm irr}(a-b)\delta . 
\label{2-m3}
\end{eqnarray}

When $H_a>H_{\rm irr}$, the magnitude of $m_y$ is reduced below that given in 
Eq.~(\ref{2-m3}), because the current-carrying regions near the edges become 
narrower. However, a more detailed theory beyond the scope of the present 
approach would be required to calculate $m_y$ for $H_a>H_{\rm irr}$. 

\subsection{Two strips in a descending field} 
\subsubsection*{Step (iv) for $H_4<H_{a\downarrow}<H_{\rm irr}$} 
If the applied field $H_a$ has been above $H_{\rm irr}$ and now decreases 
through $H_{\rm irr}$, magnetic flux is expelled from the slit and penetrates 
into the strips from the inner edge at $x=\pm b$, because the inner-edge field 
$H_y(b_-,0)$ is equal to the flux-entry field $H_s$. Magnetic flux in the 
strips escapes from the outer edges at $x=\pm a$. The domelike flux 
distributions at $\beta_4<|x|<a_-$ shrink as $H_a$ decreases 
[Fig.~\ref{Fig_HJ-2}(d)]. Current spikes occur near $x\simeq\pm a$ in 
Fig.~\ref{Fig_HJ-2}(d) because a finite current $K_z\neq 0$ flows in the 
vicinity of the outer edges, $a_-<|x|<a$, whereas $K_z=0$ at 
$\beta_4<|x|<a_-$. The parameters are given by $\alpha=a_-$ and 
$\beta=\beta_4$, where $\beta_4$ is determined by $H_y(b_-,0)=H_s$, 
\begin{eqnarray}
   \beta_4^2 &=& b_-^2 +\left(\frac{H_s}{H_a}\right)^2 
     \frac{f_2(b_-)}{a_-^2-b_-^2} 
\nonumber\\
   &\simeq& b^2+2\delta b[(H_s/H_a)^2-1] . 
\label{2-b4}
\end{eqnarray}
The magnetic moment $m_y$ is given by 
\begin{eqnarray}
   m_y/\pi &=& -H_a (a^2+b^2-a_-^2-\beta_4^2) 
\nonumber\\
   &\simeq& 2(H_s^2/H_a)b\delta -2H_a(a+b)\delta . 
\label{2-m4}
\end{eqnarray}

Step (iv) terminates and the domelike field distributions in the strips 
disappear when $\beta_4=a_-$ at $H_a=H_4$ [Fig.~\ref{Fig_HJ-2}(e)], where 
\begin{equation}
   \frac{H_4}{H_s}
   = \frac{\sqrt{f_2(b_-)}}{a_-^2-b_-^2} 
   \simeq \sqrt{\frac{2b\delta}{a^2-b^2}} . 
\label{2-H4}
\end{equation}

\subsubsection*{Step (v) for $H_5<H_{a\downarrow}<H_4$} 
During this step, positive magnetic flux exits from the slit, penetrates 
into the strips from the inner edges at $x=\pm b$, flows outward entirely 
across the strips, and annihilates at the outer edges ($x=\pm a$), where 
$H_y(a_+,0)<0$. The field distribution is shown in Fig.~\ref{Fig_HJ-2}(f). 

The parameter $\alpha=\beta=\alpha_5$ is determined by $H_y(b_-,0)=H_s$, 
\begin{eqnarray}
   \alpha_5^2&=& b_-^2 +(H_s/H_a)\sqrt{f_2(b_-)} 
\nonumber\\
   &\simeq& b^2 +(H_s/H_a)\sqrt{2\delta b(a^2-b^2)} . 
\label{2-a5}
\end{eqnarray}
Note that $\alpha_5$ is real (and $\alpha_5>a_-$) for $0<H_a<H_4$, that 
$\alpha_5\to\infty$ for $H_a\to +0$, and that $\alpha_5$ is imaginary 
(i.e., $\alpha_5^2<0$) for $H_5<H_a<0$. The magnetic moment $m_y$ is given by 
\begin{eqnarray}
   m_y/\pi &=& -H_a(a^2+b^2-2\alpha_5^2) 
\nonumber\\
   &\simeq& -H_a(a^2-b^2) +2H_s\sqrt{2\delta b(a^2-b^2)} . 
\label{2-m5}
\end{eqnarray}
Even when $H_a<0$, positive magnetic flux is still trapped in the slit, 
and the magnetic moment is positive. 

Step (v) terminates when $H_y(a_+,0)=-H_s$ at $H_a=H_5<0$, 
\begin{eqnarray}
   \frac{-H_5}{H_s} &=& \frac{1}{a_+^2-b_-^2} 
     \left[ \sqrt{f_2(a_+)} -\sqrt{f_2(b_-)} \right] 
\nonumber\\
   &\simeq& \left(\sqrt{a}-\sqrt{b}\right)\sqrt{\frac{2\delta}{a^2-b^2}} . 
\label{2-H5}
\end{eqnarray}

\subsubsection*{Step (vi) for $-H_1<H_{a\downarrow}<H_5$} 
During this step, negative magnetic flux (i.e., flux lines aligned along the 
$-y$ axis) penetrates into the strips from the outer edges ($x=\pm a$), flows 
inward entirely across the strips, and annihilates at the inner edges 
($x=\pm b$), where $H_y(b_-,0)>0$. The parameter $\alpha=\beta=\alpha_6$ is 
determined by $H_y(a_+,0)=-H_s$, and is given by $\alpha_6(H_a) 
=\alpha_1(-H_a)$, where $\alpha_1(H_a)$ is defined in Eq.~(\ref{2-a1}). 
Note that $\alpha_6$ is imaginary (i.e., $\alpha_6^2<0$) for $H^*<H_a<H_5$, 
and that $0<\alpha_6<b_+$ for $-H_1<H_a<H^*$, where $H_1$ is defined in 
Eq.~(\ref{2-H1}) and 
\begin{equation}
   \frac{-H^*}{H_s}= \frac{\sqrt{f_2(a_+)}}{a_+^2} 
   \simeq \sqrt{\frac{2(a^2-b^2)\delta}{a^3}} . 
\label{2-H*}
\end{equation}
The magnetic field at the center of the slit, ${\cal H}(0)=H_y(0,0)= 
H_a\alpha\beta/ab$, obeys ${\cal H}(0)>0$ for $H_{a\downarrow}>H^*$, 
and ${\cal H}(0)<0$ for $H_{a\downarrow}<H^*$. 

\subsubsection*{Behavior for $H_{a\downarrow}<-H_1$} 
The magnetic response of two strips for $H_{a\downarrow}<-H_1$ in descending 
fields is very similar to the response for $H_{a\uparrow}>H_1$ in ascending 
fields. The complex field ${\cal H}(\zeta,H_a)$ and magnetic moment 
$m_y(H_a)$ in descending field can be determined from the ascending-field 
results with the help of the symmetries ${\cal H}(\zeta,H_{a\downarrow}) 
=-{\cal H}(\zeta,-H_{a\uparrow})$ and $m_y(H_{a\downarrow}) 
=-m_y(-H_{a\uparrow})$, respectively.

\subsection{Magnetization curves of two strips} 
\begin{figure}
\includegraphics{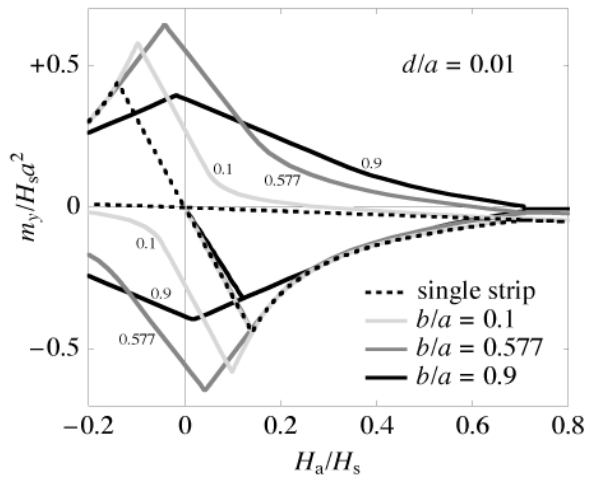}
\caption{%
Hysteretic behavior of the magnetic moment $m_y$ of two strips as a function 
of the applied magnetic field $H_a$. Magnetization curves of a single strip 
(dashed lines) and two strips (solid lines) for $b/a=0.1$, $0.577$, and $0.9$. 
Thickness of the strips is $d/a=0.01$.}
\label{Fig_MH-2}
\end{figure}

Figure~\ref{Fig_MH-2} shows hysteretic $m_y$ versus $H_a$ curves for two 
strips. In ascending fields $H_{a\uparrow}$, the sheet current $K_z(x)$ 
concentrates near the outer edges of the strips at $x\simeq\pm a$ 
[Figs.~\ref{Fig_HJ-2}(b) and (c)], and therefore, even for large $b/a$, 
the $m_y$ for two strips is almost the same as that for a single strip 
except at low fields, $H_a/H_s\alt 0.2$. 

In descending magnetic fields $H_{a\downarrow}$, we see in Fig.~\ref{Fig_MH-2} 
a striking difference between $m_y>0$ for two strips and $m_y\approx 0$ 
(i.e., $0<-m_y\ll H_s a^2$) for a single strip. The inner edges of the two 
strips at $x=\pm b$ are responsible for the large positive magnetic moment. 
The edge barriers near $x=\pm b$ prevent magnetic-flux penetration from the 
slit region $|x|<b$ into the superconducting region $b<|x|<a$. Magnetic flux 
is therefore trapped in the slit as shown in Figs.~\ref{Fig_HJ-2}(d)-(f), 
and the magnetic moment becomes positive ($m_y>0$). 

The remanent magnetic moment $m_{\rm rem}$ is given by Eq.~(\ref{2-m5}) with 
$H_a=0$, 
\begin{equation}
   m_{\rm rem}/\pi\simeq 2H_s\sqrt{2\delta b(a^2-b^2)} , 
\label{2-m_rem}
\end{equation}
which is maximized when $b/a=1/\sqrt{3}\approx 0.577$. If the slit is made 
wider, $m_{\rm rem}$ increases as $b/a$ increases for $0<b/a<1/\sqrt{3}$, 
because the slit can trap a larger amount of magnetic flux. If the slit is too 
wide, however, $m_{\rm rem}$ decreases for $1/\sqrt{3}<b/a<1$, because the 
superconducting strips (i.e., the current-carrying region) become too narrow. 
A similar behavior occurs for the peak of the magnetic moment, $m_{\rm peak}$, 
  which occurs at $H_a=H_5$. Equations~(\ref{2-m5}) and (\ref{2-H5}) yield 
\begin{equation}
   m_{\rm peak}/\pi\simeq 
     H_s \left(\sqrt{a}+\sqrt{b}\right)\sqrt{2(a^2-b^2)\delta} , 
\label{2-m_peak}
\end{equation}
which is maximized when $b/a\approx 0.403$. 

\section{\label{3_strips}%
Three strips (strips with two slits)} 
\subsection{Complex field for three strips} 
In this section we consider three strips (i.e., strips with two slits) of 
total width $2a$, as shown in Fig.~\ref{Fig_strips}(c). The outer strips are 
at $b<|x|<a$, the inner strip is at $|x|<c$, and the slits are at $c<|x|<b$. 
The strips are connected at the ends, $|z|\to\infty$, so that a circulating 
current flows in the outer strips, $\int_{-a}^{-b}K_z dx 
=-\int_{+b}^{+a}K_zdx\neq 0$. The inner strip carries no net current, 
$\int_{-c}^{+c}K_zdx=0$. The general expressions for the complex field and 
the magnetic moment per unit length for three strips are 
\begin{eqnarray}
   {\cal H}(\zeta )&=& H_a\sqrt{\frac{(\zeta^2-\alpha^2) 
     (\zeta^2-\beta^2)(\zeta^2-\gamma^2)}{%
     (\zeta^2-a^2)(\zeta^2-b^2)(\zeta^2-c^2)}} , 
\label{3-H}\\
   m_y &=& -\pi H_a\left( a^2+b^2+c^2 
     -\alpha^2-\beta^2-\gamma^2 \right) . 
\label{3-my}
\end{eqnarray}

In this section we use parameters, $a_{\pm}=a\pm\delta$, $b_{\pm}=b\pm\delta$, 
and $c_{\pm}=c\pm\delta$, where $\delta\sim\max(d,\Lambda)$. We also 
define a function $f_3(x)=(x^2-a^2)(x^2-b^2)(x^2-c^2)$ for convenience. 

\begin{figure*}
\includegraphics{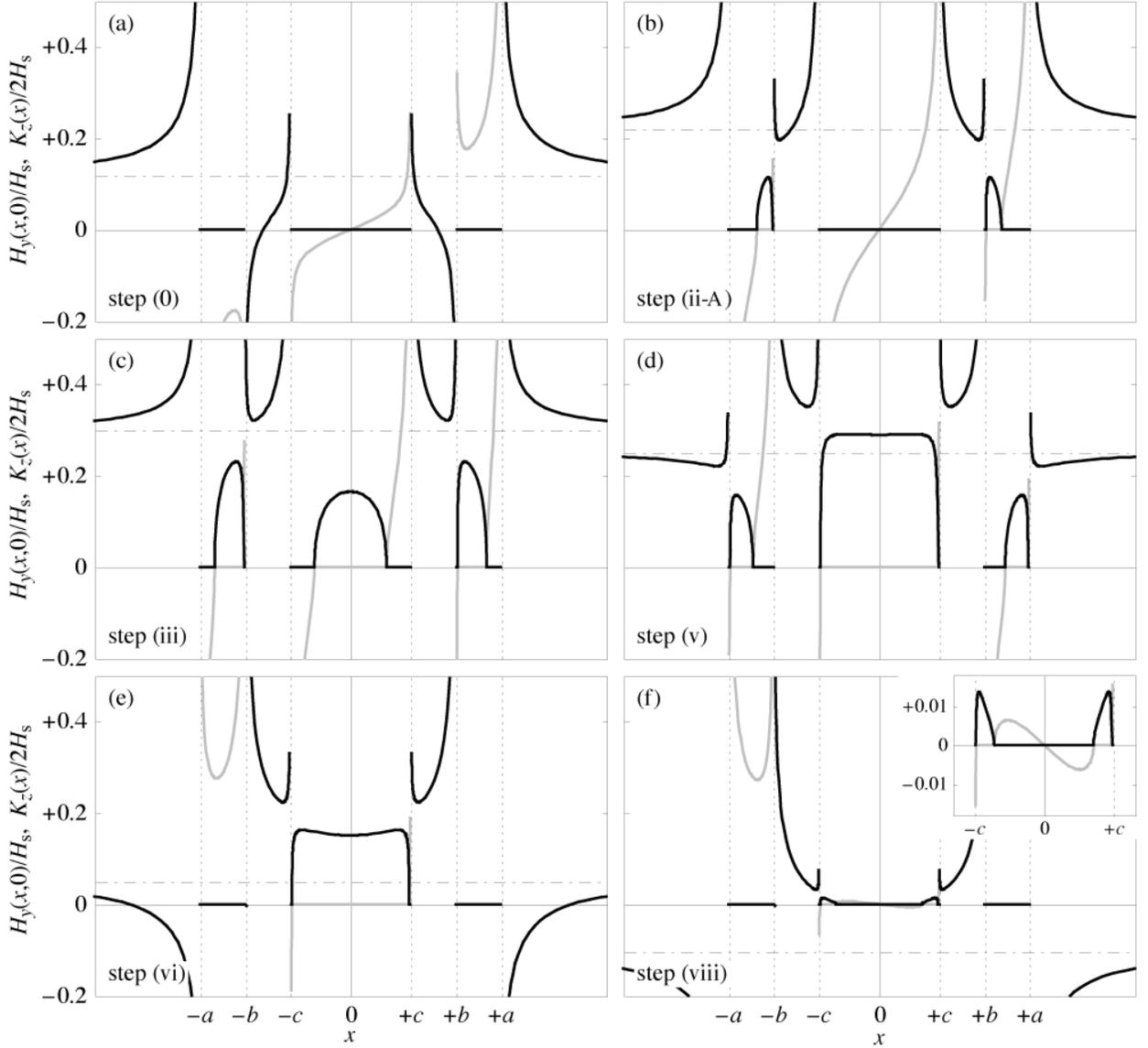}
\caption{%
Field distributions of $H_y(x,0)$, shown as black lines, and $K_z(x)/2 
=\mp H_x(x,\pm 0)$, shown as gray lines, in three strips for which $b/a=0.7$ 
and $c/a=0.4$: (a) $0<H_{a\uparrow}<H_0$, (b) $H_{\rm 1A}<H_{a\uparrow} 
<H_{\rm 2A}$, (c) $H_{\rm 2A}<H_{a\uparrow}<H_{\rm irr}$, 
(d) $H_5<H_{a\downarrow}<H_{\rm irr}$, (e) $H_6<H_{a\downarrow}<H_5$, 
and (f) $H_8<H_{a\downarrow}<H_7$ (magnified distribution in the inset). 
The horizontal dot-dashed lines show the applied magnetic field $H_a$.}
\label{Fig_HJ-3}
\end{figure*}

\subsection{Three strips in an ascending field} 
\subsubsection*{Step (0) for $0<H_{a\uparrow}<H_0$} 
When a magnetic field $H_a$ is applied after zero-field cooling [step (0)], 
the superconducting strips are initially in the Meissner state. 
Figure~\ref{Fig_HJ-3}(a) shows the distributions of $H_y(x,0)$ and 
$K_z(x)/2=\mp H_x(x,\pm 0)$. The parameters in Eq.~(\ref{3-H}) are 
$\alpha=\beta=\alpha_0$ and $\gamma=0$, where $c<\alpha_0<b$. Because the net 
magnetic flux in the slits is zero, $\int_c^b H_y(x,0)dx=0$, we find that 
\begin{equation}
   \alpha_0^2= a^2 -(a^2-c^2)\frac{E(k)}{K(k)} , \quad
   k=\sqrt{\frac{b^2-c^2}{a^2-c^2}} . 
\label{3-a0}
\end{equation}
The magnetic moment is 
\begin{equation}
   m_y/\pi= -H_a(a^2+b^2+c^2-2\alpha_0^2) . 
\label{3-m0}
\end{equation}

Step (0) terminates when $H_y(a_+,0)=H_s$ at $H_a=H_0$, where 
\begin{equation}
   \frac{H_0}{H_s}= \frac{\sqrt{f_3(a_+)}}{a_+(a_+^2-\alpha_0^2)} 
   \simeq \frac{\sqrt{2\delta a(a^2-b^2)(a^2-c^2)}}{a(a^2-\alpha_0^2)} . 
\label{3-H0}
\end{equation}

\subsubsection*{Step (i) for $H_0<H_{a\uparrow}< 
(H_{\rm 1A}\mbox{ or }H_{\rm 1B})$}
In this step, magnetic flux nucleates at the outer edges, $x=\pm a$, flows 
inward across the outer strips, and penetrates into the slits; that is, 
$H_y(a_+,0)=H_s$ and $\int_c^b H_y(x,0)dx>0$. However, no magnetic flux 
penetrates into the inner strip because $H_y(c_+,0)<H_s$. The parameters 
are given by $\alpha=\beta=\alpha_1$ and $\gamma=0$, where $c<\alpha_1<b_+$. 
The value of $\alpha_1$ is determined from $H_y(a_+,0)=H_s$, 
\begin{eqnarray}
   \alpha_1^2 &=& a_+^2 -(H_s/H_a) \sqrt{f_3(a_+)}/a_+ 
\nonumber\\
   &\simeq& a^2 -(H_s/H_a) \sqrt{2(a^2-b^2)(a^2-c^2)\delta/a} . 
\label{3-a1}
\end{eqnarray}
The magnetic moment $m_y$ is given by 
\begin{eqnarray}
   m_y/\pi &=& -H_a (a^2+b^2+c^2-2\alpha_1^2) 
\nonumber\\
   &\simeq& +H_a(a^2-b^2-c^2) 
\nonumber\\
   && -2H_s\sqrt{2(a^2-b^2)(a^2-c^2)\delta/a} . 
\label{3-m1}
\end{eqnarray}
The sign of $dm_y/dH_a=\pi(a^2-b^2-c^2)$ can be either positive or negative, 
depending upon the relative widths of the strips. 

Step (i) terminates when either $\alpha_1=b_+$ at $H_a=H_{\rm 1A}$ or 
$H_y(c_+,0)=H_s$ at $H_a=H_{\rm 1B}$. The characteristic fields $H_{\rm 1A}$ 
and $H_{\rm 1B}$ are given by 
\begin{eqnarray}
   \frac{H_{\rm 1A}}{H_s} &=& \frac{\sqrt{f_3(a_+)}}{a_+(a_+^2-b_+^2)} 
   \simeq \sqrt{\frac{2(a^2-c^2)\delta}{a(a^2-b^2)}} , 
\label{3-H1A}\\
   \frac{H_{\rm 1B}}{H_s} &=& \frac{1}{a_+^2-c_+^2} 
     \left[\ \frac{\sqrt{f_3(a_+)}}{a_+} 
     +\frac{\sqrt{f_3(c_+)}}{c_+} \ \right] 
\nonumber\\
   &\simeq& \sqrt{\frac{2\delta}{a^2-c^2}} 
     \left(\sqrt{\frac{a^2-b^2}{a}} +\sqrt{\frac{b^2-c^2}{c}}\right) . 
\label{3-H1B}
\end{eqnarray}
When $b^2<a^2-ac+c^2$, step (i) terminates at $H_a=H_{\rm 1A}<H_{\rm 1B}$, 
whereas when $b^2>a^2-ac+c^2$, step (i) terminates at $H_a=H_{\rm 1B} 
<H_{\rm 1A}$. 

\subsubsection*{Step (ii-A) for $H_{\rm 1A}<H_{a\uparrow}<H_{\rm 2A}$ 
and $b^2<a^2-ac+c^2$} 
During this step, domelike distributions of magnetic flux are present at 
$b_+<|x|<\alpha_{\rm 2A}$ in the outer strips, whereas no magnetic flux 
penetrates into the inner strip [Fig.~\ref{Fig_HJ-3}(b)]. The parameters are 
given by $\alpha=\alpha_{\rm 2A}$, $\beta=b_+$, and $\gamma=0$, where 
$b_+<\alpha_{2A}<a_-$, and $\alpha_{\rm 2A}$ is determined by 
$H_y(a_+,0)=H_s$, 
\begin{eqnarray}
   \alpha_{\rm 2A}^2 &=& a_+^2 -\left(\frac{H_s}{H_a}\right)^2 
     \frac{f_3(a_+)}{a_+^2(a_+^2-b_+^2)} 
\nonumber\\
   &\simeq& a^2 -2(H_s/H_a)^2 (a^2-c^2)\delta/a . 
\label{3-a2A}
\end{eqnarray}
The magnetic moment $m_y$ is given by 
\begin{eqnarray}
   m_y/\pi &=& -H_a (a^2+b^2+c^2-\alpha_{\rm 2A}^2-b_+^2) 
\nonumber\\
   &\simeq& -H_ac^2 -2(H_s^2/H_a) (a^2-c^2)\delta/a . 
\label{3-m2A}
\end{eqnarray}

Step (ii-A) terminates when $H_y(c_+,0)=H_s$ at $H_a=H_{\rm 2A}$, where 
\begin{eqnarray}
   \frac{H_{\rm 2A}}{H_s} &=&
     \sqrt{\frac{1}{a_+^2-c_+^2} 
     \left[\frac{f_3(a_+)}{a_+^2(a_+^2-b_+^2)} 
     +\frac{f_3(c_+)}{c_+^2(b_+^2-c_+^2)}\right]} 
\nonumber\\
   &\simeq& \sqrt{2(a+c)\delta/ac} . 
\label{3-H2A}
\end{eqnarray}

\subsubsection*{Step (ii-B) for $H_{\rm 1B}<H_{a\uparrow}<H_{\rm 2B}$ 
and $b^2>a^2-ac+c^2$} 
In this step, a domelike distribution of magnetic flux is present for 
$|x|<\gamma_{\rm 2B}$ in the inner strip, whereas no magnetic flux is present 
in the outer strips. The parameters are given by $\alpha=\beta=\alpha_{\rm 2B}$ 
and $\gamma=\gamma_{\rm 2B}$, where $0<\gamma_{\rm 2B}<c_-$, $c< 
\alpha_{\rm 2B}<b$, and $\alpha_{\rm 2B}$ and $\gamma_{\rm 2B}$ are determined 
by the coupled equations $H_y(a_+,0)=H_s$ and $H_y(c_+,0)=H_s$, which yield 
\begin{eqnarray}
   H_s/H_a &=& (a_+^2-\alpha_{\rm 2B}^2) 
     \sqrt{(a_+^2-\gamma_{\rm 2B}^2)/f_3(a_+)} 
\nonumber\\
   &=& (\alpha_{\rm 2B}^2-c_+^2) 
     \sqrt{(c_+^2-\gamma_{\rm 2B}^2)/f_3(c_+)} . 
\label{3-ac2B}
\end{eqnarray}
The magnetic moment $m_y$ is given by 
\begin{equation}
   m_y/\pi = -H_a (a^2+b^2+c^2-2 \alpha_{\rm 2B}^2-\gamma_{\rm 2B}^2), 
\label{3-m2B}
\end{equation}
where $\alpha_{\rm 2B}$ and $\gamma_{\rm 2B}$ must be determined numerically 
from Eq.\ (\ref{3-ac2B}). 

Step (ii-B) terminates when $\alpha_{\rm 2B}=b_+$ at $H_a=H_{\rm 2B}$, where 
\begin{eqnarray}
    \frac{H_{\rm 2B}}{H_s} 
    &=& \sqrt{\frac{1}{a_+^2-c_+^2} 
    \left[\frac{f_3(a_+)}{(a_+^2-b_+^2)^2} 
     -\frac{f_3(c_+)}{(b_+^2-c_+^2)^2} \right]} 
\nonumber\\
   &\simeq& \sqrt{\frac{2(a+c)(b^2-ac)\delta}{%
     (a^2-b^2)(b^2-c^2)}} . 
\label{3-H2B}
\end{eqnarray}

\subsubsection*{Step (iii) for $(H_{\rm 2A}\mbox{ or } 
H_{\rm 2B})<H_{a\uparrow}<H_{\rm irr}$} 
During this step, domelike distributions of magnetic flux are present both 
for $|x|<\gamma_3$ in the inner strip and $b_+<|x|<\alpha_3$ in the outer 
strips [Fig.~\ref{Fig_HJ-3}(c)]. The parameters are given by $\alpha=\alpha_3$, 
$\beta=b_+$, and $\gamma=\gamma_3$, where $0<\gamma_3<c_-$, $b_+<\alpha_3<a_-$, 
and $\alpha_3$ and $\gamma_3$ are determined by the coupled equations 
$H_y(a_+,0)=H_s$ and $H_y(c_+,0)=H_s$, which yield 
\begin{eqnarray}
   H_s/H_a &=& \sqrt{(a_+^2-\alpha_3^2) 
     (a_+^2-\gamma_3^2)(a_+^2-b_+^2)/f_3(a_+)} 
\nonumber\\
   &=& \sqrt{(\alpha_3^2-c_+^2) 
     (c_+^2-\gamma_3^2)(b_+^2-c_+^2)/f_3(c_+)} . 
\nonumber\\
\label{3-ac3}
\end{eqnarray}
The magnetic moment is given by 
\begin{eqnarray}
   m_y/\pi &=& -H_a (a^2+b^2+c^2-\alpha_3^2-b_+^2-\gamma_3^2) 
\nonumber\\
   &\simeq& 2H_a(a+b+c)\delta-2(H_s^2/H_a)(a+c)\delta . 
\label{3-m3}
\end{eqnarray}

Step (iii) terminates when $\alpha_3\simeq a_-$ and $\gamma_3\simeq c_-$ 
at $H_a=H_{\rm irr}\simeq H_s/\sqrt{2}$.

\subsubsection*{Step (iv) for $H_a>H_{\rm irr}$} 
The magnetic response is reversible for $H_a>H_{\rm irr}$. 
At $H_a=H_{\rm irr}$ the parameters are $\alpha=a_-$, $\beta=b_+$, 
and $\gamma=c_-$, and the magnetic fields at the edges are $H_y(a_+,0) 
\simeq H_y(b_-,0)\simeq H_y(c_+,0)\simeq H_s$. The magnetic moment at 
$H_a=H_{\rm irr}$ is given by 
\begin{eqnarray}
   m_y/\pi &=& -H_{\rm irr} (a^2+b^2+c^2-a_-^2-b_+^2-c_-^2) 
\nonumber\\
   &\simeq& -2H_{\rm irr}(a-b+c)\delta . 
\label{3-m4}
\end{eqnarray}

\subsection{Three strips in a descending field} 
\subsubsection*{Step (v) for $H_5<H_{a\downarrow}<H_{\rm irr}$} 
During this step, domelike distributions of magnetic flux are present for 
$\beta_5<|x|<a_-$ in the outer strips and for $|x|<c_-$ in the inner strip 
[Fig.~\ref{Fig_HJ-3}(d)]. These domelike flux distributions shrink as $H_a$ 
decreases. Magnetic flux escapes from the inner strip and penetrates into the 
slits. In turn, magnetic flux exits from the slits, penetrates into the outer 
strips, flows outward along the outer strips, and finally escapes from the 
strips at the outer edges, $x=\pm a$. The parameters are given by $\alpha=a_-$, 
$\beta=\beta_5$, and $\gamma=c_-$, where $\beta_5$ is determined by 
$H_y(b_-,0)=H_s$, 
\begin{eqnarray}
   \beta_5^2 &=& b_-^2 +\left(\frac{H_s}{H_a}\right)^2 
     \frac{f_3(b_-)}{(a_-^2-b_-^2)(b_-^2-c_-^2)} 
\nonumber\\
   &\simeq& b^2+2\delta b[(H_s/H_a)^2-1] . 
\label{3-b5}
\end{eqnarray}
The magnetic moment $m_y$ is given by 
\begin{eqnarray}
   m_y/\pi &=& -H_a (a^2+b^2+c^2-a_-^2-\beta_5^2-c_-^2) 
\nonumber\\
   &\simeq& -2H_a(a+b+c)\delta +2(H_s^2/H_a) b\delta . 
\label{3-m5}
\end{eqnarray}

Step (v) terminates when $\beta_5=a_-$ at $H_a=H_5$, where 
\begin{equation}
   \frac{H_5}{H_s}= \frac{1}{a_-^2-b_-^2} 
     \sqrt{\frac{f_3(b_-)}{b_-^2-c_-^2}} 
   \simeq \sqrt{\frac{2b\delta}{a^2-b^2}} . 
\label{3-H5}
\end{equation}

\subsubsection*{Step (vi) for $H_6<H_{a\downarrow}<H_5$} 
Throughout this step, no magnetic flux remains in the outer strips, but a 
domelike distribution of magnetic flux is still present for $|x|<c_-$ in the 
inner strip [Fig.~\ref{Fig_HJ-3}(e)]. Magnetic flux at $|x|<b$ penetrates 
into the outer strips from $x=\pm b$, flows outward entirely across the outer 
strips, and annihilates at $x=\pm a$, because $H_y(a_+,0)<0$. 
The parameters are given by $\alpha=\beta=\alpha_6$ and $\gamma=c_-$, 
where $\alpha_6$ is determined by $H_y(b_-,0)=H_s$, 
\begin{eqnarray}
   \alpha_6^2 &=& b_-^2 +\frac{H_s}{H_a} 
     \sqrt{\frac{f_3(b_-)}{b_-^2-c_-^2}} 
\nonumber\\
   &\simeq& b^2+(H_s/H_a)\sqrt{2\delta b(a^2-b^2)} . 
\label{3-a6}
\end{eqnarray}
Note that $\alpha_6$ is real (and $\alpha_6>a$) for $0<H_a<H_5$, 
that $\alpha\to+\infty$ for $H_a\to +0$, and that $\alpha_6$ is imaginary 
(i.e., $\alpha_6^2<0$) for $H_6<H_a<0$. The magnetic moment $m_y$ is given by 
\begin{eqnarray}
   m_y/\pi &=& -H_a (a^2+b^2+c^2-2\alpha_6^2-c_-^2) 
\nonumber\\
   &\simeq& -H_a(a^2-b^2) +2H_s\sqrt{2\delta b(a^2-b^2)} . 
\label{3-m6}
\end{eqnarray}
In the thin-strip limit $d/a\to 0$, the remanent magnetic moment $m_{\rm 
rem}$,  given by Eq.\ (\ref{3-m6}) with $H_a=0$, is the same as for two 
strips,  Eq.\ (\ref{2-m_rem}). 

Step (vi) terminates when $H_y(a_+,0)=-H_s$ at $H_a=H_6<0$, where 
\begin{eqnarray}
   \frac{-H_6}{H_s} &=& \frac{1}{a_+^2-b_-^2} 
     \left[\ \sqrt{\frac{f_3(a_+)}{a_+^2-c_-^2}} 
     -\sqrt{\frac{f_3(b_-)}{b_-^2-c_-^2}} \ \right] 
\nonumber\\
   &\simeq& \left(\sqrt{a}-\sqrt{b}\right)\sqrt{\frac{2\delta}{a^2-b^2}} . 
\label{3-H6}
\end{eqnarray}

\subsubsection*{Step (vii) for $H_7<H_{a\downarrow}<H_6<0$} 
During this step (not shown in Fig.~4), the magnetic field at the outermost 
edges is equal to the negative flux-entry field, $H_y(a_+,0)=-H_s$. 
Negative magnetic flux penetrates into the outer strips at $x=\pm a$ and flows 
entirely across the outer strips into the slits, resulting in a reduction of 
magnetic flux in both the slits and the center strip. The parameter 
$\gamma=c_-$ is fixed, and $\alpha=\beta=\alpha_7$ is imaginary 
(i.e., $\alpha_7^2<0$), where $\alpha_7$ is determined by $H_y(a_+,0)=-H_s$, 
\begin{eqnarray}
   \alpha_7^2 &=& a_+^2 +\frac{H_s}{H_a} 
     \sqrt{\frac{f_3(a_+)}{a_+^2-c_-^2}} 
\nonumber\\
   &\simeq& a^2 +(H_s/H_a)\sqrt{2\delta a(a^2-b^2)} . 
\label{3-a7}
\end{eqnarray}
The magnetic moment is given by 
\begin{eqnarray}
   m_y/\pi &=& -H_a (a^2+b^2+c^2-2\alpha_7^2-c_-^2) 
\nonumber\\
   &\simeq& +H_a(a^2-b^2) +2H_s\sqrt{2\delta a(a^2-b^2)} . 
\label{3-m7}
\end{eqnarray}

Step (vii) terminates when $\alpha_7=0$ at $H_a=H_7<0$, where 
\begin{equation}
   \frac{-H_7}{H_s}= \frac{1}{a_+^2} \sqrt{\frac{f_3(a_+)}{a_+^2-c_-^2}} 
   \simeq \sqrt{\frac{2(a^2-b^2)\delta}{a^3}} . 
\label{3-H7}
\end{equation}
The magnetic field at the center, ${\cal H}(0)=H_y(0,0)$, is positive for 
$H_a>H_7$, but becomes zero at $H_a=H_7$. 

\subsubsection*{Step (viii) for $H_8<H_{a\downarrow}<H_7<0$} 
During this step, domelike distributions of magnetic flux are present at 
$\beta_8<|x|<c_-$ in the inner strip, whereas no magnetic flux is present 
in the inner region, $|x|<\beta_8$ [Fig.~\ref{Fig_HJ-3}(f)]. The parameters 
are given by $\alpha=0$, $\beta=\beta_8$, and $\gamma=c_-$, 
where $\beta_8$ is determined by $H_y(a_+,0)=-H_s$, 
\begin{eqnarray}
   \beta_8^2 &=& a_+^2 -\left(\frac{H_s}{H_a}\right)^2 
     \frac{f_3(a_+)}{a_+^2(a_+^2-c_-^2)} 
\nonumber\\
   &\simeq& a^2 -2(H_s/H_a)^2 (a^2-b^2)\delta/a . 
\label{3-b8}
\end{eqnarray}
The magnetic moment is given by 
\begin{eqnarray}
   m_y/\pi &=& -H_a (a^2+b^2+c^2-\beta_8^2-c_-^2) 
\nonumber\\
   &\simeq& -H_ab^2 -2(H_s^2/H_a)(a^2-b^2)\delta/a . 
\label{3-m8}
\end{eqnarray}

Step (viii) terminates when $\beta_8=c_-$ at $H_a=H_8<0$, where 
\begin{equation}
   \frac{-H_8}{H_s}= \frac{\sqrt{f_3(a_+)}}{a_+(a_+^2-c_-^2)} 
   \simeq \sqrt{\frac{2(a^2-b^2)\delta}{a(a^2-c^2)}} . 
\label{3-H8}
\end{equation}
At $H_a=H_8$, no magnetic flux is present in either the inner or outer strips. 

\subsubsection*{Step (ix) for $-H_1<H_{a\downarrow}<H_8<0$} 
During this step, no magnetic flux is present in the strips, and the parameters 
are given by $\alpha=0$ and $\beta=\gamma=\beta_9$. The parameter $\beta_9$, 
determined by $H_y(a_+,0)=-H_s$, is given by $\beta_9(H_a)=\alpha_1(-H_a)$, 
where $\alpha_1$ is defined in Eq.~(\ref{3-a1}). 

\subsubsection*{Behavior for $H_{a\downarrow}<-H_1<0$} 
The magnetic response of two strips for $H_{a\downarrow}<-H_1$ in a descending 
field is very similar to that for $H_{a\uparrow}>H_1$ in an ascending field. 
The complex field and magnetic moment as functions of the applied field can be 
determined with the help of the symmetries ${\cal H}(\zeta,H_{a\downarrow}) 
=-{\cal H}(\zeta,-H_{a\uparrow})$ and $m_y(H_{a\downarrow}) 
=-m_y(-H_{a\uparrow})$, respectively.

\subsection{Magnetization curves of three strips} 
\begin{figure}
\includegraphics{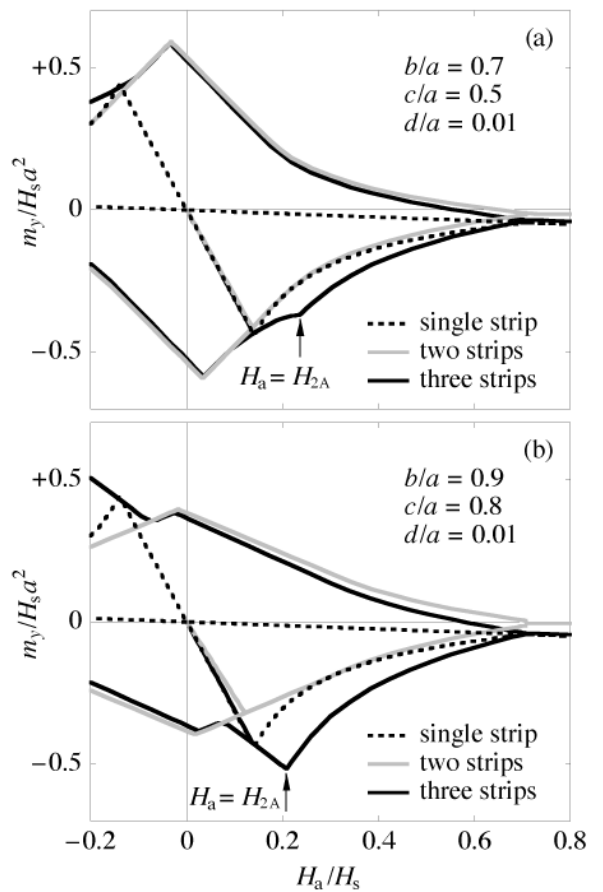}
\caption{%
Hysteretic behavior of the magnetic moment $m_y$ of three strips as a function 
of the applied magnetic field $H_a$. (a) Magnetization curves of a single 
strip (dashed lines), two strips (gray solid lines) with $b/a=0.7$, and three 
strips (black solid lines) with $b/a=0.7$ and $c/a=0.5$. (b) Magnetization 
curves as (a) but for $b/a=0.9$ and $c/a=0.8$. Thickness of the strips is 
$d/a=0.01$.}
\label{Fig_MH-3}
\end{figure}

Figure~\ref{Fig_MH-3} shows hysteretic $m_y$-$H_a$ curves for three strips. 
In ascending magnetic fields at $H_{a\uparrow}=(H_{\rm 2A}\ \mbox{or}\ 
H_{\rm 2B})$, the $m_y$-$H_a$ curves of three strips have additional kinks 
[see arrow in Fig.~\ref{Fig_MH-3}(a)] or peaks [see arrow in 
Fig.~\ref{Fig_MH-3}(b)]. Such kink or peak structures do not appear in the 
$m_y$-$H_a$ curves for a single strip or two strips. They arise in three 
strips because the the edge barrier at the edges of the center strip 
($x\simeq\pm c$) impede the entry of magnetic flux. In descending fields 
$H_{a\downarrow}$, on the other hand, the $m_y$-$H_a$ curves for three strips 
are almost the same as those for two strips; the similarity occurs because 
the edges of the center strip do not impede the exit of magnetic flux. 

In three strips, the edge barriers near $x\simeq\pm a$ and $\pm c$ are 
effective in preventing the entry of magnetic flux in $H_{a\uparrow}$, 
but are not effective in stopping the exit of magnetic flux in 
$H_{a\downarrow}$. On the other hand, while the edges at $x\simeq\pm b$ do 
not impede entering magnetic flux in $H_{a\uparrow}$, the edge barriers 
there are responsible for impeding flux exit in $H_{a\downarrow}$. 

\section{Conclusion} 
We investigated field distributions and the magnetic moment $m_y$ of 
bulk-pinning-free strips with slits in applied magnetic fields $H_a$, 
and we studied these in detail for increasing fields $H_{a\uparrow}$ 
and decreasing fields $H_{a\downarrow}$. 

For two strips in decreasing fields $H_{a\downarrow}$, the edge barriers near 
the inner edges at $x\simeq\pm b$ impede the exit of magnetic flux from the 
slit at $|x|<b$ into the strip at $b<|x|<a$. The trapping of magnetic flux in 
the slit in $H_{a\downarrow}$ results in a positive remanent magnetic moment, 
$m_y>0$ at $H_{a\downarrow}=0$, rather than a zero remanent moment, which 
occurs for a single strip without slits. For two strips, the remanent magnetic 
moment at $H_{a\downarrow}=0$ is maximized when $b/a=1/\sqrt{3}$. 

For three strips (i.e., strips at $|x|<c$ and $b<|x|<a$) the edge barriers at 
$x\simeq\pm b$ in $H_{a\downarrow}$ are effective in trapping magnetic flux as 
in two strips. In the thin-strip limit $d/a\to 0$, the remanent magnetic 
moment  for three strips at $H_{a\downarrow}=0$ is the same as for two 
strips.  In $H_{a\uparrow}$, the edge barriers at $x\simeq\pm c$ impede 
the penetration  of magnetic flux into the inner strip. As a consequence, 
the $m_y$ of three  strips in $H_{a\uparrow}$ exhibits kink or peak 
structures that are not present  for two strips. 

The above arguments for three strips can be extended to an arbitrary number of 
strips. Consider a symmetric array of $N$ coplanar strips of total width 
$2a_0$ ($N-1$ slits), in which the strip edges are at $x=\pm a_n$, where 
$a_{N-1}<a_{N-2}<\cdots<a_1<a_0$. Superconducting strips occupy the regions 
$a_{n+1}<|x|<a_n$ with $n$ even, and slits are at $a_{n+1}<|x|<a_n$ with 
$n$ odd. The region spanning the centerline, $|x| < a_{N-1}$ corresponds to 
a superconducting strip for odd $N$ and a slit for even $N$. The even edges 
at $x=\pm a_n$ for $n=0,\,2,\,4,\cdots$ impede flux entry in increasing 
applied fields $H_{a\uparrow}$, and the odd edges at $x=\pm a_{n}$ for 
$n=1,\,3,\,5,\cdots$ impede flux exit in decreasing fields $H_{a\downarrow}$. 

By considering narrow slit widths, one may further extend the above approach 
to model grain-boundary pinning in thin films. The resulting model would be 
the thin-film analog of the bulk-pinning model discussed in 
Ref.~\onlinecite{Clem74}. 

\begin{acknowledgments}
We thank H.\ Yamasaki for stimulating discussions. This work was done as a 
part of the Super-ACE project (a project for the R\&D of fundamental 
technologies for superconducting ac power equipment) of the Ministry of 
Economy, Trade, and Industry of Japan. 
This manuscript has been authored in part by Iowa State University of Science 
and Technology under Contract No.\ W-7405-ENG-82 with the U.S.\ Department of 
Energy.
\end{acknowledgments}

\end{document}